\documentclass[aps,prl,floatfix,twocolumn,tightenlines]{revtex4}
\usepackage{amsmath}
\usepackage{latexsym}
\usepackage{graphicx}
\usepackage{color}

\begin{document}

\title{Classical dispersion-cancellation interferometry}
\author{K.J. Resch$^{1,2}$, P. Puvanathasan$^{1}$, J.S. Lundeen$^{3}$, M.W.
Mitchell$^{4}$, and K. Bizheva$^{1}$} \affiliation{$^{1}$Department
of Physics \& Astronomy and $^{2}$Institute for Quantum
Computing,$\phantom{z}$University of Waterloo,$\phantom{z}$Waterloo,
ON, Canada
N2L 3G1 \\
$^{3}$Clarendon Laboratory, University of Oxford, Parks Rd, Oxford, United
Kingdom OX1 3PU \\
$^{4}$ ICFO -- Institut de Ciencies Fotoniques, Mediterranean
Technology Park, 08860, Castelldefels (Barcelona), Spain }

\begin{abstract}
\noindent Even-order dispersion cancellation, an effect previously
identified with frequency-entangled photons, is demonstrated
experimentally for the first time with a linear, classical
interferometer. A combination of a broad bandwidth laser and a high
resolution spectrometer was used to measure the intensity
correlations between anti-correlated optical frequencies. Only 14\%
broadening of the correlation signal is observed when significant
material dispersion, enough to broaden the regular interferogram by
4250\%, is introduced into one arm of the interferometer.
\end{abstract}

\maketitle

\noindent Interferometry is an indispensable tool for precision
measurements.  Low-coherence, or white-light, interferometry is used
for precise measurements of material properties, such as optical
path length and dispersion. Optical coherence tomography (OCT), a
technique for non-invasive medical imaging, is based on
low-coherence interferometry \cite{fujimoto95,fercher03}. Both white
light interferometry and OCT use broad bandwidth light sources to
achieve micrometer scale image resolution \cite{drexler04}. Although
large spectral bandwidth is essential for obtaining high resolution,
it also increases dispersive broadening of the interferogram.

Quantum metrology uses quantum mechanical features, such as
entanglement and squeezed light, to improve the sensitivity of
measurement devices \cite{giovannetti04,leibfried05}. A two-photon
quantum interferometer \cite{hong87}, based on frequency-entangled
photon pairs, has been demonstrated to be insensitive to all even
orders of dispersion (\cite{steinberg02a,steinberg02b}, also see
\cite{franson92}). This effect, known as quantum dispersion
cancellation, was proposed as the basis for \emph{quantum}-optical
coherence tomography \cite{abouraddy02} and a proof-of-principle was
demonstrated experimentally \cite{nasr03}. A very recent theoretical
model \cite{erkmen06} claims that interferometric dispersion
cancellation does not require the use of individual pairs of
entangled photons. The scheme is instead based on a nonlinear
optical interferometer that employs broad-band phase conjugation
between two reflections from the same sample. Experimental
implementation of this technique would be extremely difficult
requiring both development of novel optical sources and a suitable
method of phase conjugation. Other approaches use numerical methods
to compensate dispersion in data and images obtained with
low-coherence interferometry or OCT \cite{fercher01,
deboer01,marks03,wojtkowski04,banaszek07}.

Dispersion cancellation is staightforward in quantum interferometry,
but the methods proposed so far in classical interferometry are not.
Can we use the intuition derived from quantum technologies to
achieve dispersion cancellation in a simpler way in a classical
interferometer \cite{bennink02,ferri05,resch05}? In this work, we
show that dispersion cancellation can be achieved using only a
classical light source, linear optics, and frequency-correlated
detection. We review quantum dispersion cancellation and use its
essentials to design an analogous classical system.

Consider the nonclassical two-photon interferometer shown in Fig.1a)
\cite{hong87}. The upper path is of length, $L_1$, and the lower
path is of length, $L_2$=$L_1+\Delta$. A nonlinear crystal, pumped
by a narrow bandwidth laser of frequency $2\omega_{0}$, generates
photon pairs with central frequency $\omega_{0}$, via parametric
down-conversion into the upper and lower paths of the interferometer
in the state,
\begin{eqnarray}
\left|\psi\right> = \int d\delta\omega
A(\omega_{0}+\delta\omega)|\omega_{0}+\delta\omega\rangle_{1}|\omega_{0}-\delta\omega\rangle_{2},
\end{eqnarray}
\noindent The subscripts 1 and 2 are mode labels, and
$A(\omega_{0}+\delta\omega)$ is the amplitude for a pair of photons
of frequencies $\omega_{0}+\delta\omega$ in mode 1 and
$\omega_{0}-\delta\omega$ in mode 2. The sum of the frequencies in
each term of the superposition is fixed by energy conservation --
this is a frequency-entangled state with perfect frequency
anti-correlation.  The photons are interferometrically combined at
the 50/50 beamsplitter followed by two single-photon counting
detectors. The signal of interest is the number of coincident photon
detection events as a function of the optical delay, $\Delta$.

Insertion of a dispersive, lossless medium of length, $L$, in the
upper path of the interferometer results in a frequency-dependent
phase shift, $\phi_{M}(\omega)$=$k_{M}(\omega)L$, where
$k_{M}(\omega)$ is the wavevector at frequency, $\omega$, in the
material. We series expand $k_{M}(\omega)$ about $\omega_{0}$:
\begin{equation}
k_{M}(\omega )\approx k(\omega_{0})+\left.\frac{dk}{d\omega
}\right|_{\omega_0}\delta \omega
+\frac{1}{2}\left.\frac{d^{2}k}{d\omega
^{2}}\right|_{\omega_{0}}\delta \omega^{2}+ ... ,
\end{equation}
\noindent The first derivative is the inverse of the group velocity,
$v_{g}$, at $\omega_{0}$ and leads to a shift in the centre of the
interference pattern. The second derivative is the leading-order
dispersive term, which causes loss of both spatial resolution and
contrast in low-coherence interferometry by broadening the width and
reducing the visibility of the interference pattern.

Following Ref.\cite{steinberg02a}, we make the assumption
$A(\omega_{0}+\delta\omega)$=$A(\omega_{0}-\delta\omega)$, i.e., the
amplitude is symmetric about the central frequency $\omega_{0}$.  We
find the expected coincidence rate, as a function of $\Delta$ is:
\vspace{0mm}
\begin{eqnarray} \label{quantumcase} \hspace{-10mm}C(\Delta ) &\propto
&\int d\delta \omega \left| A(\omega
_{0}+\delta \omega )\right| ^{2} \notag\\
&&\left\{ 1-\cos \left[ \frac{2\delta \omega (L+\Delta )}{c}-2L\frac{dk}{%
d\omega }\delta \omega \right] \right\}.
\end{eqnarray}
\noindent The expression is in agreement with Ref.
\cite{steinberg02a}, but uses slightly different notation. Notice
that the second derivative does not appear -- this is the dispersion
cancellation. In fact, all higher-order even derivatives are
cancelled. Maximum destructive interference occurs when the argument
of the cosine term is zero for every frequency. This happens when
the extra group delay imposed by the material is balanced by extra
optical delay in the other arm of the interferometer. We will refer
back to this expression when describing our classical system.

Now consider the Mach-Zehnder interferometer in Fig.1b).  The
dimensions and mode labels of this interferometer are identical to
that described in Fig. 1a, as is the dimension of the dispersive
material; both beamsplitters are 50/50.  The intensity spectrum of
the input light is $I(\omega)$.  The intensities registered by the
spectrometers for a delay position, $\Delta$, and frequency,
$\omega$, in the outputs labeled $a$ and $b$ are,
\begin{eqnarray}
\label{intensity1}
I_{a}(\omega,\Delta) & = & I(\omega)\cos^2
\left[ \frac{(\Delta+L)\frac{\omega}{c}-\phi_{M}(\omega)}{2} \right] \\
\label{intensity2}
I_{b}(\omega,\Delta) & = & I(\omega)\sin^2
\left[\frac{(\Delta+L)\frac{\omega}{c}-\phi_{M}(\omega)}{2} \right].
\end{eqnarray}
Each of these intensities is affected by all orders of dispersion in
the series expansion of $\phi_{M}(\omega)$.

Quantum dispersion cancellation is a result of
frequency-entanglement in fourth-order, i.e., coincidence,
detection. Our approach seeks to mimic this effect as closely as
classical physics allows. We use frequency \emph{correlations}, the
classical analogue to entanglement, and measure a fourth-order
signal, achieved by multiplying pairs of intensity measurements.
Specifically, we measure the signal, $\mathcal{S}$,
\begin{eqnarray}  \label{signaldefinition}
\mathcal{S}(\Delta) & = & \int d\delta\omega
I_{a}(\omega_{0}+\delta\omega)I_{b}(\omega_{0}-\delta\omega)
\end{eqnarray}
\noindent The integrand of this function is the product of two
intensities with an energy sum of $2\omega_{0}$.
\begin{figure}[tbp]
\includegraphics[width=1 \columnwidth]{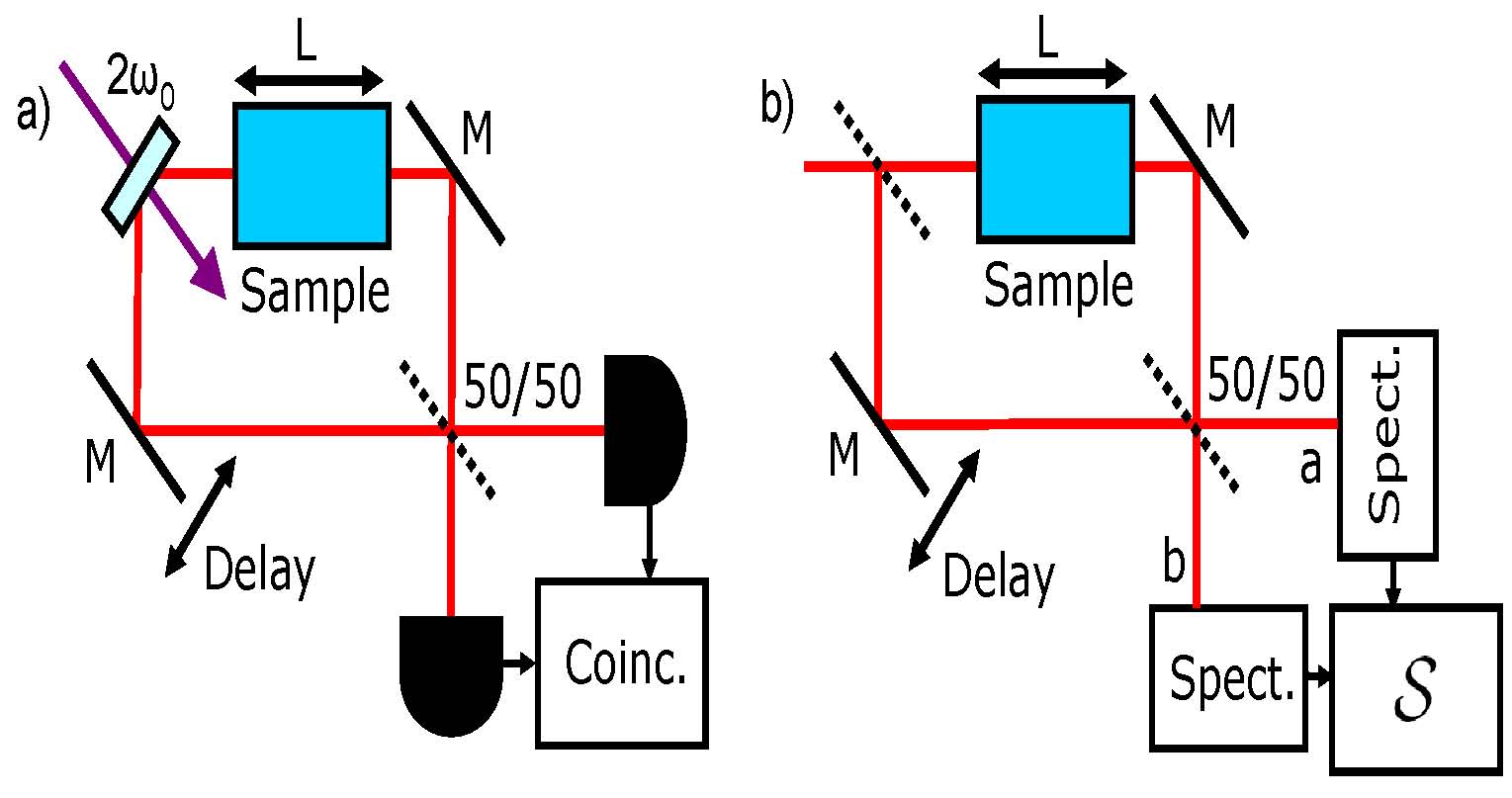}
\includegraphics[width=1 \columnwidth]{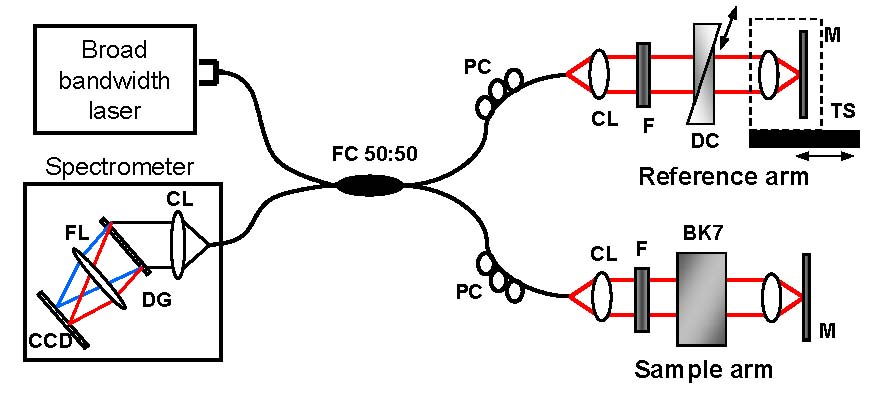}
\vspace{-0.5cm} \caption{Dispersion cancellation interferometry in a
a) Two-photon interferometer using frequency-entangled photon pairs
\cite{steinberg02a,steinberg02b} and b) a white-light Mach-Zehnder
interferometer with frequency-correlated detection. These
interferometers and their expected output signals are described in
the text.  Coinc. is coincidence detection; $\mathcal{S}$ describes
a function of the output from the spectrometers. c) Experimental
realization.  A broadband laser is the source for a fibre-based
two-path (Michelson) interferometer. The setup uses a 50/50
beamsplitter (FC 50:50), polarization controllers (PC), collimating
lenses (CL), neutral-density filters (F), two BK7 prisms for
dispersion control (DC), a translation stage (TS), mirrors (M), and
a spectrometer.  The spectrometer contains a CL, a diffraction
grating (DG), and focusing lens (FL).} \vspace{-0.4cm}
\end{figure}
We use Eqs. \ref{intensity1} \& \ref{intensity2} and assume, as we
did in the quantum case, that the input spectrum is symmetric about
$\omega_{0}$, i.e., $I(\omega_{0}+\delta)=I(\omega_{0}-\delta)$, to
obtain,
\begin{eqnarray}
\label{signal1}
    \mathcal{S}(\Delta) & = & \frac{1}{2} \int d\delta\omega
    \left[I(\omega_{0}+\delta \omega)\right]^{2}\\
    && \left\{ \begin{array}{c} 1-\frac{1}{2}\cos \left[\frac{2\delta \omega(L+\Delta)}{c}
    -2L\frac{dk}{d\omega}\delta \omega\right] \\
    -\frac{1}{2}\cos \left[\frac{2\omega_{0}(L+\Delta)}{c}+2Lk_{0}+
    L\frac{d^{2}k}{d\omega^{2}}(\delta \omega)^{2} \right] \end{array}
    \right\} \notag
\end{eqnarray}
This is the signal of interest from our classical system and can be
directly compared with the quantum signal in Eq.\ref{quantumcase}.
The argument in the first cosine term is identical to the quantum
expression and describes a dispersion-cancelled interference dip.
The second cosine term does not appear in the quantum case. Notice
though, that its argument has only weak dependence on the frequency
difference $\delta\omega$ (the integration variable) through the
dispersion term.  It describes a rapidly oscillating component, with
wavelength $\lambda_{0}=\pi c/\omega_{0}$, with a slowly decaying
envelope. The separation of length-scales between these terms allows
removal of the unwanted fast oscillation in the final data with for
example, a low-pass Fourier filter. The other feature of interest in
our classical expression is that the first term is multiplied by
1/2. This imposes the well-known classical limit of 50\% on the
destructive interference visibility in two-photon interferometers
\cite{ghosh87,hong87}.  The signal $\mathcal{S}$ is the classical
analogue to the Hong-Ou-Mandel dip \cite{hong87} and contains the
same resistance to dispersion as its quantum counterpart.

The experimental setup is shown in Fig. 2. A compact,
fiber-pigtailed, femtosecond laser (Femtolasers Inc., centre
wavelength 792nm, bandwidth FWHM 154nm, average power 60mW) was
coupled to a fiber-based Michelson interferometer. Broad bandwidth
optical and fiber optic components were chosen to support
propagation of the entire laser bandwidth with minimal spectral and
power losses. A pair of BK7 prisms mounted on miniature translation
stages in the reference arm of the system were used to precisely
compensate material dispersion mismatch between the two arms of the
interferometer. The focusing lens and the mirror in the reference
arm of the system were mounted on a computer-controlled translation
stage for variable optical delay. The interference pattern generated
by light reflected from the sample and reference mirrors was
detected with a high-resolution (0.09nm) and high-speed (20 kHz
readout rate) spectrometer and recorded by a computer.  The
spectrometer utilized a 4096 pixel linear-array CCD camera and it
was calibrated for the spectral range 607nm to 1012nm.  To
demonstrate dispersion cancellation with the classical
interferometer, measurements were made both in a dispersion-balanced
system and when flat, uncoated, BK7 optical windows of thickness
$4.690\pm0.005$mm, $5.940\pm0.005$mm, and $6.170\pm0.005 $mm (and
several possible combinations) were introduced into the sample arm.
For each measurement, the reference mirror was translated in steps
of 0.1$\mu$m and the spectral interference fringes were acquired
with a readout time of 60$\mu$s per step -- at least 4 orders of
magnitude shorter as compared with typical measurement times in
entangled photon experiments.

The calculation of the signal function $\mathcal{S}$ was performed
in the following way.  One spectrometer reading was taken for each
motor position to provide us with $I_a(\lambda,\Delta)$.  The
wavelength scale was converted to frequency and nonlinear
interpolation was used to extract intensities at evenly spaced
intervals. We obtained $I(\omega)$ by measuring the intensity from
the sample and reference arm separately and doubling their sum.
Energy conservation,
$I(\omega)$=$I_{a}(\omega,\Delta)$+$I_{b}(\omega,\Delta)$, was
applied to extract $I_{b}(\omega,\Delta)$ without the necessity for
a second spectrometer. To satisfy the assumption in our theory that
$I(\omega)$=$I(2\omega_{0}-\omega)$, $I(\omega)$ and $I_{a}(\omega)$
was multiplied by a mirror version of $I(\omega)$ with respect to
the centre frequency $\omega_{0}$. The integral $\mathcal{S}$ was
approximated by a discrete sum over 4096 equally spaced energies.
\begin{figure}[tbp]
\vspace{-0mm}\centerline{
    \mbox{\includegraphics[width=0.5 \columnwidth]{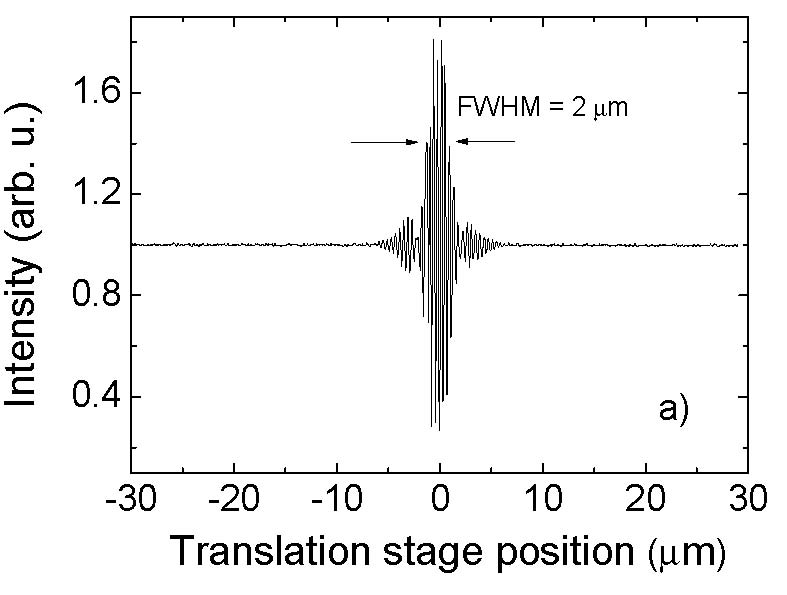}}
    \mbox{\includegraphics[width=0.48 \columnwidth]{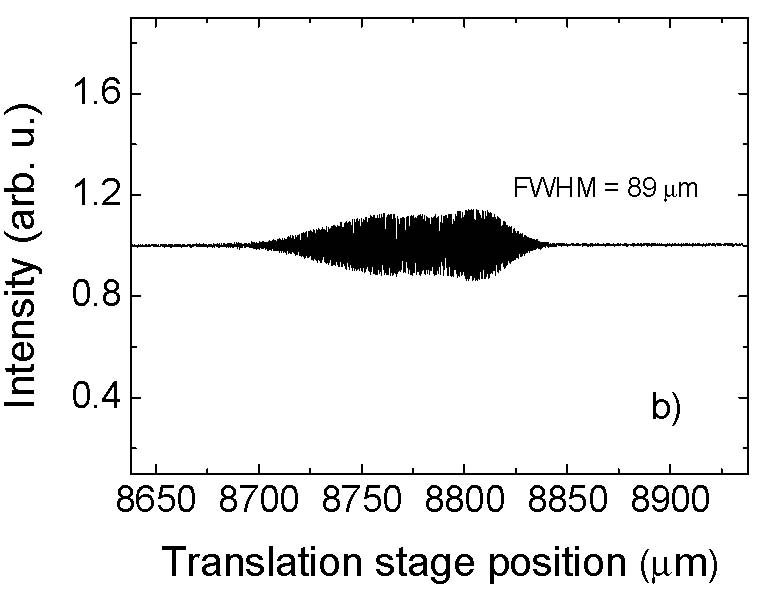}}
  }
\vspace{-0mm}\centerline{
    \mbox{\includegraphics[width=0.5 \columnwidth]{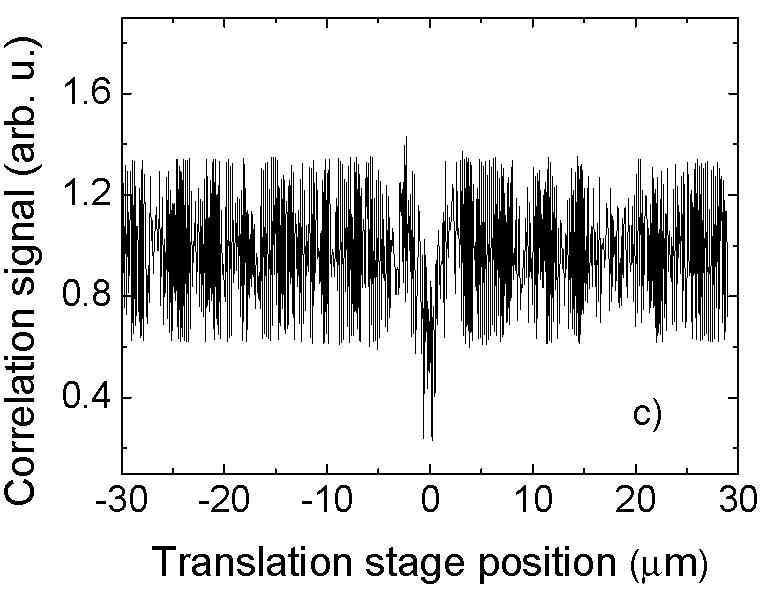}}
    \mbox{\includegraphics[width=0.48 \columnwidth]{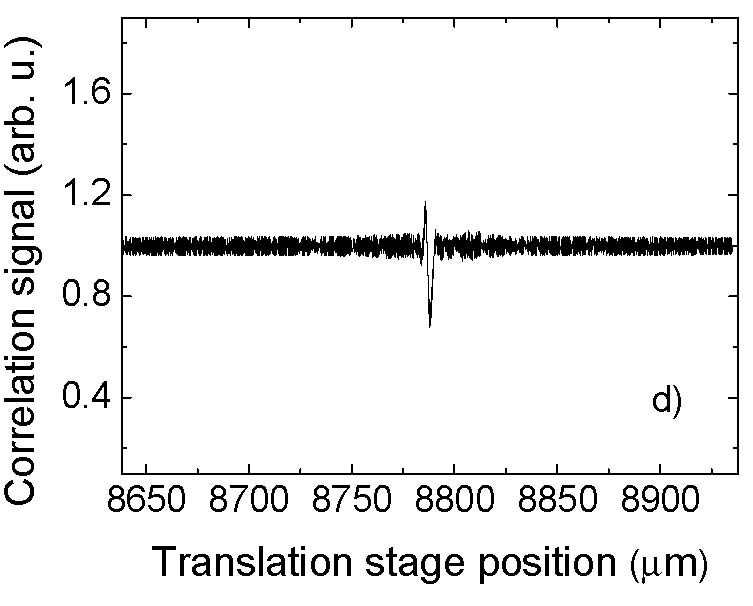}}
  }
\vspace{-0mm} \centerline{
    \mbox{\includegraphics[width=0.5 \columnwidth]{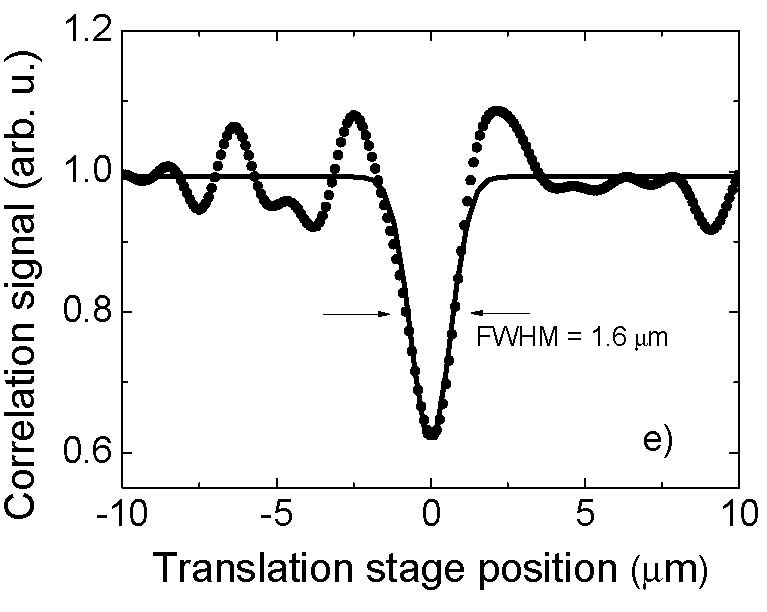}}
    \mbox{\includegraphics[width=0.5 \columnwidth]{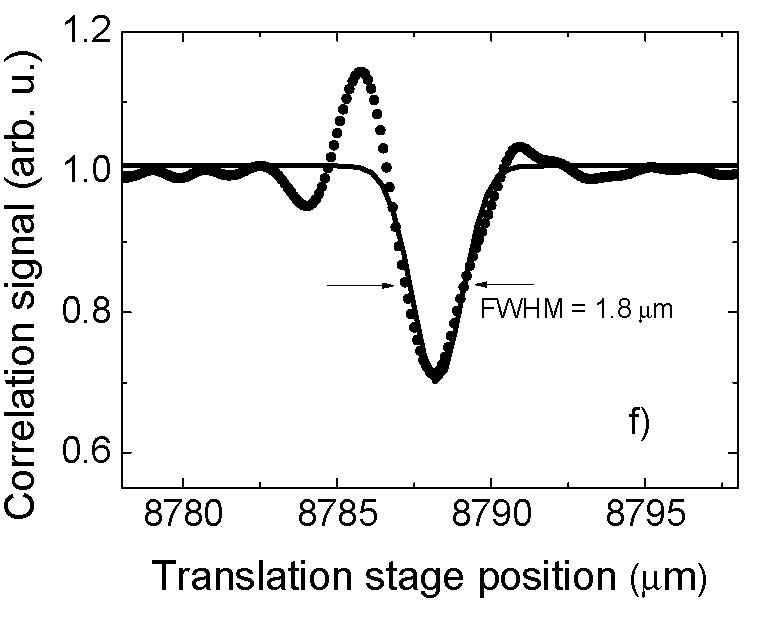}}
  }
\vspace{-0.3cm} \caption{Experimental Data.  a) \& b) Total
intensity, as measured by summing the intensities measured at each
frequency by the spectrometer, versus motor position with $0$ and
two passes through $(16.800\pm0.009)$mm of BK7 glass in the sample
arm of the interferometer, respectively. c) \& d) $\mathcal{S}$
versus motor position with no BK7 and 16.8mm of BK7 in the sample
arm.  In e) \& f), the data from c) \& d) have been subject to a
Fourier low-pass filter to remove rapidly oscillating terms.  The
solid curves are Gaussian fits.  These data show that $\mathcal{S}$
broadens by only about 14\% by addition of the glass while the
standard intensity interference pattern is broadened by 4250\%.
}\vspace{-0.4cm}
\end{figure}

The total intensity registered by the spectrometer was obtained by
adding the intensity measured at each pixel at a fixed translation
stage position. This signal is equivalent to a signal that one would
have been measured by a square-law detector, such as a photo-diode.
Two examples of the total intensity measured as a function of the
translation stage position are shown in Fig.2 for the cases where no
glass (a) and a $16.800\pm0.009$mm thick BK7 glass window (b) were
inserted into the sample arm of the interferometer. As a result of
the material dispersion, the intensity interference pattern is
dramatically broadened, from $(2.04\pm0.03)$$\mu$m to
$(88.6\pm0.9)$$\mu$m, and the fringe visibility is reduced
\cite{visdefinition}, from 78\% to 14\%. This figure clearly shows
the detrimental effect that dispersion has on interference.

The corresponding correlation signal function, $\mathcal{S}$, for
the two cases of no material dispersion and $16.800$mm BK7 glass are
shown in Figs. 2c) and 2d), respectively. Each of these signals has
a sharp dip in addition to a rapidly oscillating component that
corresponds to the final cosine term in Eq.\ref{signal1}. Note that
the magnitude of the fast oscillating signal is \emph{reduced} when
a large amount of dispersion is present in the interferometer. A
similar effect was observed when we simulated the measurements with
a computer model. The data from Fig. 2c) and 2d) was filtered with a
low-pass Fourier filter to remove the fast oscillating term and the
the filtered data is presented in Figs. e) and f). These dips were
fitted to a Gaussian function to extract their centres and FWHM.
While the intensity interference pattern is broadened by 4250\% of
its original size due to material dispersion, the correlation signal
$\mathcal{S}$ is broadened only by 14\%. The visibilities of the
correlation signal dip \cite{visdefinition} is reduced from
$(40.8\pm0.14)\%$ and $(30.0\pm0.3)\%$ for Figs. e) and f),
respectively (recall that the theoretical maximum visibility is
50\%). The deviations from the Gaussian shape of the fitting
function are due partially to the non-Gaussian spectrum of the laser
as well as the present of higher-order material dispersion.
\begin{figure}[tbp]
\centerline{
    \mbox{\includegraphics[width=0.50 \columnwidth]{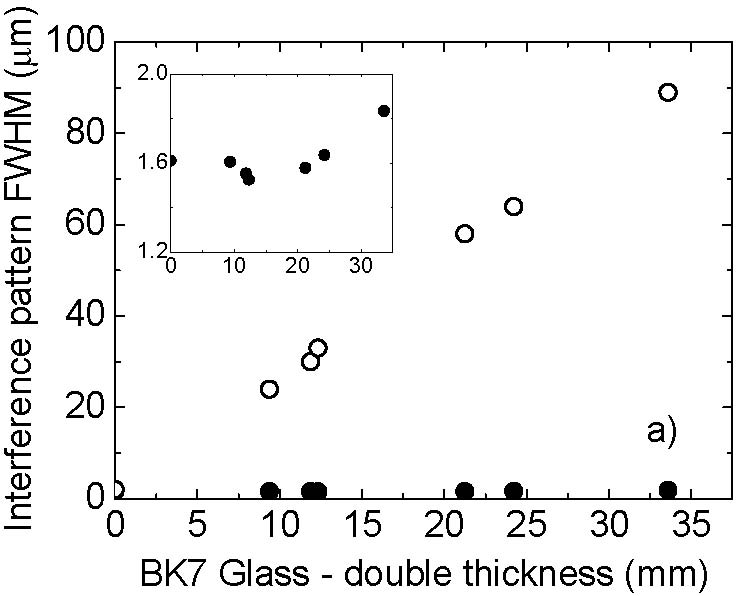}}
    \mbox{\includegraphics[width=0.49 \columnwidth]{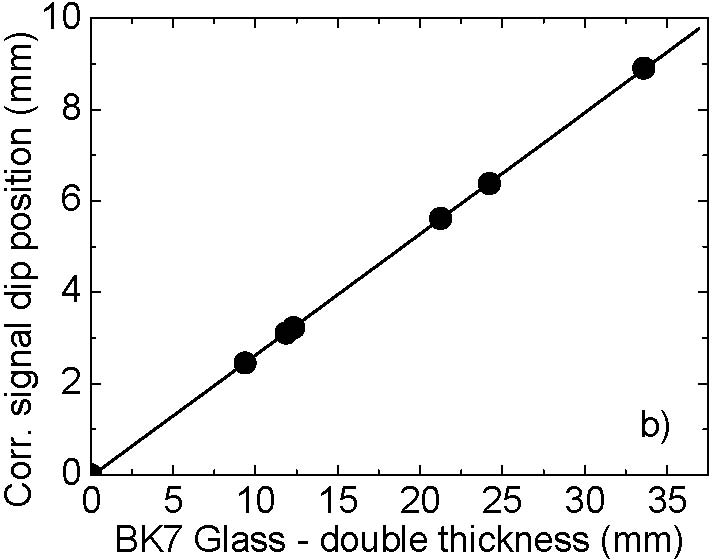}}
  }
\vspace{-0.1cm}\caption{Performance of classical dispersion
cancellation. a) The width of the interference patterns (total
intensity, open circle; $\mathcal{S}$, solid circle), as measured by
translation stage displacement, versus the thickness of the glass
traversed by the beam, i.e., twice the glass thickness.  The inset
expands the y-axis to show the almost constant width of
$\mathcal{S}$ over the whole range of glass thicknesses.  b) The
shift in the centre of the interference pattern versus the thickness
of the glass. As discussed in the text, these data show that the
interference pattern is displaced by the group delay.}
\vspace{-0.4cm}
\end{figure}

Fig. 3a) is a plot of the total intensity FWHM (open circles) and
the correlation signal $\mathcal{S}$ FWHM (solid circles) as
functions of twice the physical thickness of the BK7 optical flats
(we use double the thickness because the Michelson interferometer
uses a reflection geometry).

The relative shift in the correlation signal dip centre as a
function of twice the physical thickness of the glass in presented
in Fig 3b). We estimated the statistical error in the centre of the
dip to be about 1$\mu$m based on the standard deviation of 5
consecutive measurements.  This is about a factor of 50-100 times
larger than the fitting error and could be improved by using a
higher precision translation stage. However, the most significant
measurement error is associated with the widths of the BK7 flats. We
expect the shift in the centre of the dip to be determined by the
group velocity by the relation, $(c_{\rm{air}}/v_{g}-1)L$, where
$c_{\rm{air}}$ is the speed of light in air and $v_{g}$ is the group
velocity, in our case at a wavelength of 800nm. From Fig 3b), we
evaluate the slope $0.2633$. The accepted value, $0.2631$, was
obtained from the Sellmeier equation \cite{ghosh97}.  Our errors are
dominated by the uncertainty in the material thickness, which is
about 0.1\%, and at this level the slope from the data and the
theory agree. The centre of the correlation signal $\mathcal{S}$ is
determined by the group velocity.

We have theoretically derived and experimentally demonstrated a
method for cancelling even-order dispersion in classical
low-coherence interferometry. Dispersion cancellation is not a
uniquely quantum effect, since it can also be observed in completely
classical systems.  However, the interference visibility in our
classical analogue is only half that achievable in quantum
interferometers \cite{hong87,steinberg02b}.  Two seemingly
contradictory constraints are essential in both the quantum and
classical techniques: a wide bandwidth of frequencies provides good
time resolution, whereas narrow frequency correlations reduce
sensitivity to dispersion.  Our approach dramatically reduces
experimental barriers for dispersion cancellation in low-coherence
interferometry and optical coherence tomography.

We thank Aephraim Steinberg for valuable discussions and Gregor
Weihs for loaning us equipment. This work was funded by the
University of Waterloo, NSERC (KB), and OPC (KB).


\begin{thebibliography}{99}

\bibitem{fujimoto95} J.G. Fujimoto et al., \emph{Nat. Med.} \textbf{1}, 970
(1995).

\bibitem{fercher03} A.F. Fercher et al., \emph{Rep. Prog. Phys.} \textbf{66}, 239
(2003).

\bibitem{drexler04} W. Drexler, J. Biomed. Opt. \textbf{9}, 47 (2004).

\bibitem{giovannetti04} V. Giovannetti et al.,
\emph{Science} \textbf{306}, 1330 (2004).

\bibitem{leibfried05} Leibfried D et al. \emph{Nature} \textbf{438}, 639
(2005).

\bibitem{hong87}C.K. Hong et al., \emph{Phys. Rev. Lett.} \textbf{59},
2044 (1987).

\bibitem{steinberg02a} A.M. Steinberg et al.,
\emph{Phys. Rev. A} \textbf{45}, 6659 (1992).

\bibitem{steinberg02b} A.M. Steinberg et al.,
\emph{Phys. Rev. Lett.} \textbf{68}, 2421 (1992).

\bibitem{franson92} J.D. Franson, \emph{Phys. Rev. A} \textbf{45}, 3126
(1992).

\bibitem{abouraddy02} A. F. Abouraddy, et al., \emph{Phys. Rev. A} \textbf{65}, 053817
(2002).

\bibitem{nasr03} M. B. Nasr, et al., \emph{Phys. Rev. Lett.} \textbf{91}, 083601 (2003).

\bibitem{erkmen06} B.I. Erkmen and J.H. Shapiro, \emph{Phys. Rev. A} \textbf{%
74}, 041601, 2006

\bibitem{fercher01} A.F. Fercher et al., \emph{Optics Express},
\textbf{9}, 610 (2001).

\bibitem{deboer01} J.F. de Boer et al., \emph{Appl.
Opt.} \textbf{40}, 5787 (2001).

\bibitem{marks03} D.L. Marks et al., \emph{Appl. Opt.}, \textbf{42},
3038 (2003).

\bibitem{wojtkowski04} M. Wojtkowski et al., \emph{Optics Express}, \textbf{12},
2404 (2004).

\bibitem{banaszek07} K. Banaszek et al.,
\emph{Opt. Comm.} \textbf{269}, 152 (2007).

\bibitem{bennink02} R. S. Bennink et al.,
\emph{Phys. Rev. Lett.} \textbf{89}, 113601 (2002).

\bibitem{ferri05} F. Ferri et al., \emph{Phys. Rev. Lett.} \textbf{94}, 183602
(2005).

\bibitem{resch05} K.J. Resch et al. \emph{Time-reversal and
super-resolving phase measurements} quant-ph/0511214 (2005).

\bibitem{ghosh87} R. Ghosh and L. Mandel, \emph{Phys. Rev. Lett.} \textbf{59}, 1903
(1987).

\bibitem{visdefinition} Visibility is a measure of the contrast of
interference.  For oscillating interference, visibility is
$V_{osc}$=$(I_{Max}-I_{Min})/(I_{Max}+I_{Min})$, where $I_{Max}$ and
$I_{Min}$ are the maximum and minimum of the pattern. For
interference dips the visibility is conventionally
$V_{dip}$=$(I_{Max}-I_{Min})/I_{Max}$.

\bibitem{ghosh97} G. Ghosh, \emph{Appl. Opt.} \textbf{36}, 1540 (1997).

\end{thebibliography}
\end{document}